\documentclass[aps,prx,twocolumn,superscriptaddress,floatfix]{revtex4-2}

\usepackage{graphicx}
\usepackage{dcolumn}

\usepackage{bm}
\usepackage{amssymb}
\usepackage{amsmath}
\usepackage{microtype}
\usepackage{xfrac}
\usepackage{array}
\usepackage[dvipsnames]{xcolor}
\usepackage[colorlinks,plainpages=false,linkcolor=blue,urlcolor=blue,citecolor=blue,pdfpagemode=UseNone,pdfstartview=FitBH]{hyperref}
\usepackage{nicematrix}
\def\equationautorefname~#1\null{(#1)\null}

\begin{document}

\title{\texorpdfstring{$\mu$}~SR Study of the Dipole-Octupole Quantum Spin Ice Candidate \texorpdfstring{Ce$_2$Zr$_2$O$_7$}~}

\author{J.~Beare}
\email{jamesbeare.physics@gmail.com}
\affiliation{Department of Physics and Astronomy, McMaster University, Hamilton, Ontario L8S 4M1, Canada}

\author{E.~M.~Smith}
\email{J. Beare and E. M. Smith contributed equally to this work.}
\affiliation{Department of Physics and Astronomy, McMaster University, Hamilton, Ontario L8S 4M1, Canada}

\author{J.~Dudemaine}
\affiliation{D\'epartement de Physique, Universit\'e de Montr\'eal, Montr\'eal, Quebec H2V 0B3, Canada}
\affiliation{Regroupement Qu\'eb\'ecois sur les Mat\'eriaux de Pointe (RQMP), Quebec H3T 3J7, Canada}

\author{R.~Sch\"{a}fer}
\affiliation{Department of Physics, Boston University, Boston, Massachusetts 02215, USA}

\author{M.~R.~Rutherford}
\affiliation{Department of Physics and Astronomy, McMaster University, Hamilton, Ontario L8S 4M1, Canada}

\author{S.~Sharma}
\affiliation{Department of Physics and Astronomy, McMaster University, Hamilton, Ontario L8S 4M1, Canada}

\author{A.~Fitterman}
\affiliation{D\'epartement de Physique, Universit\'e de Montr\'eal, Montr\'eal, Quebec H2V 0B3, Canada}
\affiliation{Regroupement Qu\'eb\'ecois sur les Mat\'eriaux de Pointe (RQMP), Quebec H3T 3J7, Canada}

\author{C.~A.~Marjerrison}
\affiliation{Brockhouse Institute for Materials Research, McMaster University, Hamilton, Ontario L8S 4M1, Canada}

\author{T.~J.~Williams}
\affiliation{Neutron Scattering Division, Oak Ridge National Laboratory, Oak Ridge, Tennessee 37831, USA}

\author{A.~A.~Aczel}
\affiliation{Neutron Scattering Division, Oak Ridge National Laboratory, Oak Ridge, Tennessee 37831, USA}

\author{S.~R.~Dunsiger}
\affiliation{Department of Physics, Simon Fraser University, Burnaby, British Columbia, Canada V5A 1S6}
\affiliation{TRIUMF, Vancouver, British Columbia, Canada V6T 2A3}

\author{A.~D.~Bianchi}
\affiliation{D\'epartement de Physique, Universit\'e de Montr\'eal, Montr\'eal, Quebec H2V 0B3, Canada}
\affiliation{Regroupement Qu\'eb\'ecois sur les Mat\'eriaux de Pointe (RQMP), Quebec H3T 3J7, Canada}

\author{B.~D.~Gaulin}
\affiliation{Department of Physics and Astronomy, McMaster University, Hamilton, Ontario L8S 4M1, Canada}
\affiliation{Brockhouse Institute for Materials Research, McMaster University, Hamilton, Ontario L8S 4M1, Canada}
\affiliation{Canadian Institute for Advanced Research, 661 University Avenue, Toronto, Ontario M5G 1M1, Canada.}

\author{G.~M.~Luke}
\affiliation{Department of Physics and Astronomy, McMaster University, Hamilton, Ontario L8S 4M1, Canada}
\affiliation{Brockhouse Institute for Materials Research, McMaster University, Hamilton, Ontario L8S 4M1, Canada}
\affiliation{TRIUMF, Vancouver, British Columbia, Canada V6T 2A3}

\date{\today}

\begin{abstract}
 The Ce$^{3+}$ pseudospin-$1/2$ degrees of freedom in Ce$_2$Zr$_2$O$_7$ possess both dipolar and octupolar character which enables the possibility of novel quantum spin liquid ground states in this material. Here we report new muon spin relaxation and rotation ($\mu$SR) measurements on single crystal samples of Ce$_2$Zr$_2$O$_7$ in zero magnetic field and in magnetic fields directed along the $[1,\bar{1},0]$ and $[1,1,1]$ crystallographic directions, and for magnetic fields directed both longitudinal and transverse to the direction of muon polarization. Our zero-field results show no signs of magnetic ordering or spin freezing, consistent with earlier zero-field $\mu$SR measurements on a powder sample of Ce$_2$Zr$_2$O$_7$, and also with the expectations for a quantum spin ice. However, we measure a more gentle relaxation rate for Ce$_2$Zr$_2$O$_7$ in zero-field at low temperatures than was previously reported. This difference in relaxation rate is likely due to the low oxidation, and correspondingly, the high stoichiometry of our single crystal samples. Longitudinal field measurements confirm that the magnetic dipole moments in Ce$_2$Zr$_2$O$_7$ remain dynamic at $T = 0.1$~K on the $\mu$s time scale. For both $[1,\bar{1},0]$ and $[1,1,1]$ magnetic fields, our $\mu$SR Knight shift measurements show a field-induced leveling off of the magnetic susceptibility at low temperature which is qualitatively consistent with corresponding calculations using the numerical-linked-cluster method in combination with recent estimates for the nearest-neighbour exchange parameters of Ce$_2$Zr$_2$O$_7$. 
\end{abstract}

\maketitle

\section{\label{sec:Intro}Introduction}

\begin{figure}[t]
\linespread{1}
\par
\includegraphics[width=3.4in]{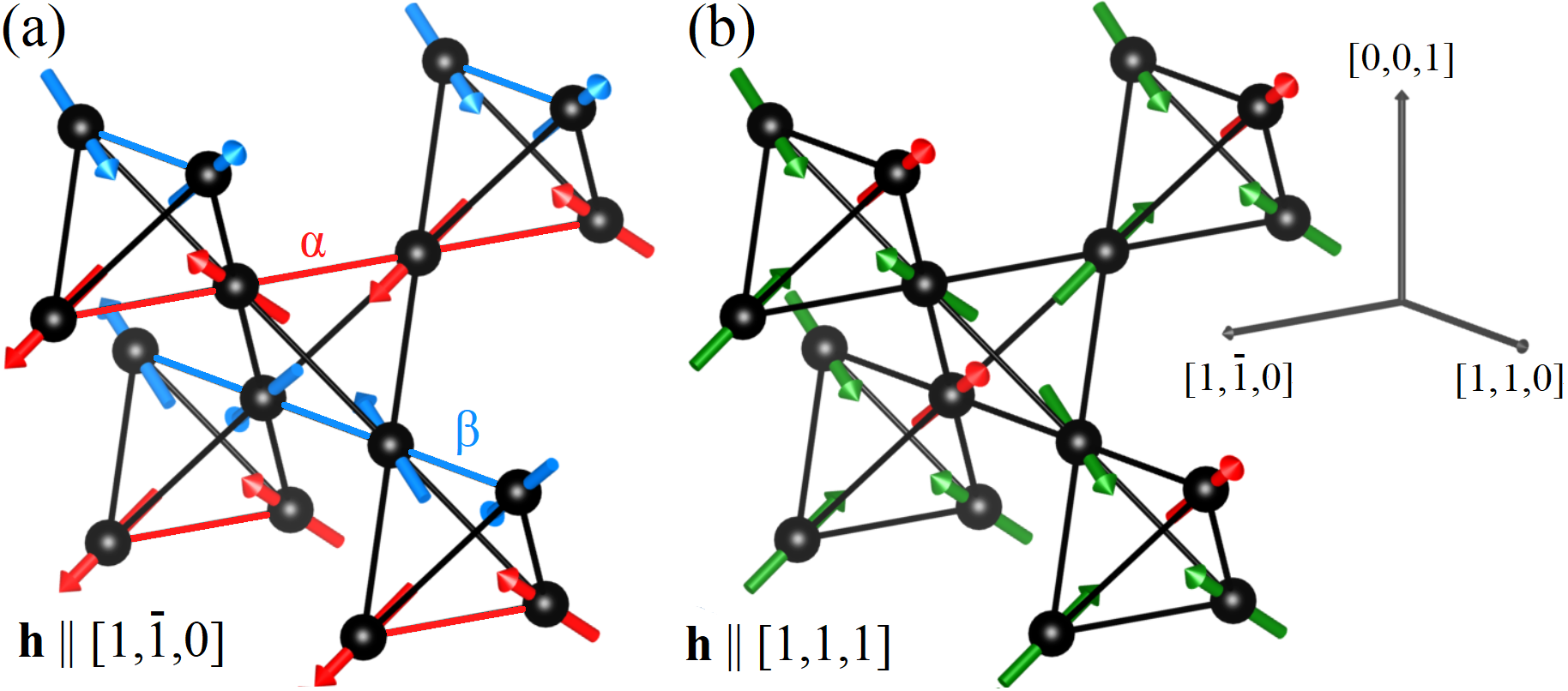}
\par
\caption{The magnetic rare-earth ions composing five corner-sharing tetrahedra within the pyrochlore crystal structure. (a) The local magnetic structure expected in a moderate-strength $[1,\bar{1},0]$ magnetic field for spin ice with ferromagnetic coupling of dipole moments, showing the decomposition of the magnetic sublattice into $\alpha$ chains (red) and $\beta$ chains (blue). The $\alpha$ chains (red) are along the field direction and are field-polarized to the extent allowable by the Ising single ion anisotropy, and the $\beta$ chains (blue) are perpendicular to the field direction and show short-ranged ferromagnetic order with two possible, energetically-equivalent directions for the order. The short-ranged ferromagnetic order of the $\beta$ chains locally establishes the two-in, two-out spin ice rule for dipole moments. (b) The three-in, one-out magnetic structure expected for Ising pyrochlores in a moderate-strength magnetic field along the $[1,1,1]$ direction, where the magnetic moments along the field direction are shown in red and the magnetic moments of the other rare-earth ions, which form Kagome planes perpendicular to the $[1,1,1]$ magnetic field, are shown in green.} 
\label{Fig:1}
\end{figure}

Quantum spin liquid phases have been ``cause-c\'el\`ebres" in condensed matter physics for much of the last 50 years, as they arise in the context of quantum magnetism and geometrical frustration, and they are expected to support low temperature, quantum disordered states with exotic excitations. Unlike conventional magnetic phases, quantum spin liquid phases display neither magnetic order nor spin freezing down to zero temperature, and can be described within quantum gauge field theories wherein their elementary excitations take the form of emergent quasiparticles that have no counterpart in conventionally ordered magnets~\cite{Balents2010}. Quantum spin ice (QSI) is a particular class of quantum spin liquid that can exist in different flavours on the three-dimensional pyrochlore lattice, a cubic network of corner-sharing tetrahedra [part of which is shown in Figs.~\ref{Fig:1}{\color{blue}(a,b)}]. Many such pyrochlores have the chemical composition $A_2B_2$O$_7$, where the $A$-site is occupied by a trivalent rare-earth ion and the $B$-site is occupied by a tetravalent transition metal ion. Theoretically, QSIs are known to possess an emergent quantum electrodynamics whose low-energy excitations are described as gapless photons and two sets of gapped excitations: electric monopoles (also known as visons), and magnetic monopoles (also known as spinons) \cite{Benton2012, GingrasReview2014, Chen2023}. 

Conventional spin ice physics is associated with magnetic moments that display a local Ising anisotropy such that they are constrained to point into or out-of the tetrahedra on which they reside, along with a ferromagnetic-coupling of these moments at the nearest-neighbour level. Even classically, such systems are known to possess a macroscopic degeneracy that precludes the formation of an ordered state at low temperatures. This disordered magnetic state is formally similar to that expected for proton disorder in hexagonal water-ice, hence the origin of the name ``spin ice". Within the disordered ground state, spins on a given tetrahedron will obey ``ice rules" such that the four spins on each tetrahedra take on local two-in, two-out configurations, wherein two magnetic moments at the corner of each tetrahedron point towards the centre of the tetrahedron, while the other two point directly away from the centre [Fig.~\ref{Fig:1}{\color{blue}(a)}]~\cite{Harris1997, Ramirez1999, Bramwell2001, Fennell2009, Morris2009, Bramwell2001b, Rau2015}. 

More generally the rare-earth pyrochlore oxides display a wealth of both exotic and conventional magnetic phases, which depend on the anisotropies describing the moments, and the details of the interactions between them~\cite{Balents2010, GingrasReview2014, GardnerReview2010, Hallas2018, RauReview2019}. The rare-earth ions display strong spin-orbit interactions that dominate over the crystal electric field (CEF) interactions in these materials, and both of these tend to dominate the weaker-still intersite exchange interactions, giving rise to a convenient separation of energy scales~\cite{Hallas2018, RauReview2019}. So long as the CEF ground state doublet is well separated from the excited states, as is the case for Ce$_2$Zr$_2$O$_7$ where the separation between the ground state doublet and the first excited CEF state doublet is $\sim$55~meV, the low temperature behaviour of such systems can be described by interacting pseudospin-$1/2$ degrees of freedom~\cite{Hallas2018, RauReview2019}. 


Three 4\textit{f}$^1$ Ce$^{3+}$ pyrochlore oxides have been synthesized and studied to date: Ce$_2$Zr$_2$O$_7$, Ce$_2$Hf$_2$O$_7$, and Ce$_2$Sn$_2$O$_7$~\cite{Gaudet2019, Gao2019, Sibille2015, Sibille2020, Smith2022, Gao2022, Yahne2022, Poree2023a, Poree2023b, Smith2023}. They have attracted much recent attention as it has been recognized that the pseudospin-$1/2$ degrees of freedom associated with Ce$^{3+}$ in these materials have both dipolar and octupolar character due to their dipole-octupole (DO) CEF ground state doublet wavefunctions~\cite{Gaudet2019, Gao2019, Sibille2015,Sibille2020,Poree2022}. The DO nature of the pseudospin-$1/2$ degrees of freedom is a consequence of how the pseudospins transform under the point group symmetry at the Ce$^{3+}$ site and under time reversal symmetry. The $x$ and $y$ components of the pseudospin-$1/2$ degrees of freedom carry octupolar magnetic moments, while the $z$ component is associated with a magnetic dipole moment~\cite{RauReview2019, Patri2020}. However, both the $x$ and $z$ component of the pseudospin-$1/2$ degrees of freedom {\it transform} like dipoles under the point group symmetries of the rare earth site in the pyrochlore lattice and time reversal symmetry, while only the $y$-component of each pseudospin-$1/2$ degree of freedom transforms like an octupole~\cite{RauReview2019, Li2017, Huang2014}. Dipole-octupole symmetry is accompanied by an Ising single-ion anisotropy for each magnetic ion due to the fact that the magnetic dipole moment of each ion comes solely from the $z$ component of pseudospin.

Importantly, the form of the microscopic spin Hamiltonian for such DO doublets derives from these same symmetry properties and can be written at the nearest-neighbour level as~\cite{Huang2014,Li2017,RauReview2019,Benton2020}:

\begin{equation}\label{eq:1}
\begin{split}
    \mathcal{H}_\mathrm{DO} & = \sum_{\langle ij \rangle}[J_{x}{S_i}^{x}{S_j}^{x} + J_{y}{S_i}^{y}{S_j}^{y} + J_{z}{S_i}^{z}{S_j}^{z} \\ 
    & + J_{xz}({S_i}^{x}{S_j}^{z} + {S_i}^{z}{S_j}^{x})] - g_z \mu_\mathrm{B} \sum_{i} (\mathbf{h} \cdot \hat{{\bf z}}_i) {S_i}^{z} \;\;.
\end{split}
\end{equation}

\noindent where ${S_{i}}^{\alpha}$ ($\alpha = x$, $y$, $z$) are the pseudospin-$1/2$ components of rare-earth atom $i$ in the local $\{x, y, z\}$ coordinate frame, the magnetic field is denoted as $\mathbf{h}$, and $\hat{{\bf z}}_i$ is the local $z$ axis for ion $i$~\cite{Huang2014, Li2017}. A rotation of the local coordinate systems about the $y$-axes, by an amount $\theta = (1/2)\tan^{-1}[2J_{xz}/(J_{z}-J_{x})]$, then yields the XYZ Hamiltonian~\cite{Huang2014,Li2017,Benton2020},

\begin{equation}\label{eq:2}
\begin{split}
    \mathcal{H}_\mathrm{XYZ} & = \sum_{\langle ij \rangle}[     J_{\tilde{x}}{S_i}^{\tilde{x}}{S_j}^{\tilde{x}} + J_{\tilde{y}}{S_i}^{\tilde{y}}{S_j}^{\tilde{y}} + J_{\tilde{z}}{S_i}^{\tilde{z}}{S_j}^{\tilde{z}}] \\ 
    & - g_z \mu_\mathrm{B} \sum_{i} \mathbf{h} \cdot\hat{{\bf z}}_i({S_i}^{\tilde{z}}\cos\theta + {S_i}^{\tilde{x}}\sin\theta) \;\;.
\end{split}
\end{equation}

This spin Hamiltonian on the pyrochlore lattice [Eq.~\autoref{eq:2}] has been well studied theoretically, and its zero-field magnetic ground state phase diagram is known to display at least four different flavours of QSI ground state, as well as two ``all-in, all-out'' N\'eel states, such that both the quantum disordered and N\'eel ordered states can possess either a dipolar or an octupolar character~\cite{Benton2020, Patri2020, Huang2018b, Kim2022, Desrochers2022}. 

In recent works, Smith \emph{et al.} (Ref.~\cite{Smith2022, Smith2023}) present a detailed case for an exotic quantum spin ice phase at low temperature in the cerium-based pyrochlore oxide Ce$_2$Zr$_2$O$_7$, and identify this phase as a novel quantum spin ice near the boundary between the dipolar and octupolar regimes. Independent work by Bhardwaj \emph{et al.} (Ref.~\cite{Changlani2022}) compares a different set of experimental results to different but related calculations from those used in Refs.~\cite{Smith2022, Smith2023}, and conclude that the magnetic ground state in Ce$_2$Zr$_2$O$_7$ is firmly within the octupolar quantum spin ice regime. A dipolar (octupolar) spin ice has the $\tilde{x}$ or $\tilde{z}$ (has $\tilde{y}$) components of pseudospin, with dipolar (octupolar) symmetry, associated with the emergent electric field~\cite{Huang2014, Li2017, Benton2020, Patri2020, Huang2018b, Kim2022, Desrochers2022}. Each of these quantum spin ices can be obtained by adding quantum fluctuations to a classical phase that is governed by a ``two-plus, two-minus'' rule for the component of pseudospin associated with the dominant exchange parameter and emergent electric field~\cite{Huang2014, Li2017, Benton2020, Patri2020, Huang2018b, Kim2022, Desrochers2022}, analogous to the conventional two-in, two-out rule for the $z$ components of pseudospin in conventional spin ices~\cite{Banerjee2008, Benton2012, Shannon2012, GingrasReview2014, Huang2014, Savary2017}. While the experimentally estimated exchange parameters vary slightly between Ref.~\cite{Changlani2022} and Ref.~\cite{Smith2022}, both works point towards a new quantum spin ice phase in Ce$_2$Zr$_2$O$_7$ with the zero-field ground state rule governing the behaviour of octupolar magnetic moments. A quantum spin ice ground state is also consistent with the original reports (Refs.~\cite{Gaudet2019, Gao2019}) on the magnetic behaviour in Ce$_2$Zr$_2$O$_7$, detailing a lack of both magnetic order and spin freezing, accompanied by a snowflake-like pattern of diffuse neutron scattering that is characteristic of QSI correlations. 

Similarly, experiments on powder samples of the dipole-octupole pyrochlore Ce$_2$Sn$_2$O$_7$ at low temperature have been interpreted in terms of an octupole-based QSI phase~\cite{Sibille2015, Sibille2020, Poree2023a}. However, recent results on hydrothermally-grown powder and single crystal samples of Ce$_2$Sn$_2$O$_7$ suggest that the ground state in these new crystals of Ce$_2$Sn$_2$O$_7$ is an all-in, all-out ordered phase that is proximate in phase space to a quantum spin ice phase whose dynamics persist down to very low temperature~\cite{Yahne2022}. 

Finally, recent work on a third cerium-based, dipole-octupole pyrochlore, Ce$_2$Hf$_2$O$_7$~\cite{Poree2022}, presented analysis of a collection of experimental data that was fit to constrain the nearest-neighbour exchange parameters. This work concluded that the corresponding ground state in Ce$_2$Hf$_2$O$_7$ is a quantum spin ice, and one whose exchange interactions are dominated by either $J_x$ or $J_y$~\cite{Poree2023b}. 

In this work, we focus on Ce$_2$Zr$_2$O$_7$ in zero field and in magnetic fields along the $[1,\bar{1},0]$ and $[1,1,1]$ directions. The effects of a $[1,\bar{1},0]$ magnetic field on the low temperature magnetic phase in spin ice materials has been investigated in a multitude of previous works, including those on Ce$_2$Zr$_2$O$_7$ (Ref.~\cite{Smith2023}), as well as on Ho$_2$Ti$_2$O$_7$ (Refs.~\cite{Harris1997, Melko2004, Fennell2005, Clancy2009}) and Dy$_2$Ti$_2$O$_7$ (Refs.~\cite{Yoshida2004, Fennell2002, Melko2004, Fennell2005, Ruff2005}) which are both disordered classical spin ices at low temperature in zero magnetic field~\cite{Harris1997, Ramirez1999, Bramwell2001, Fennell2009, Morris2009}. 

For Ising pyrochlores in a $[1,\bar{1},0]$ magnetic field, it is convenient to decompose the magnetic sublattice into two sets of orthogonal chains: so-called $\alpha$ chains that are parallel to the $[1,\bar{1},0]$ field direction and so-called $\beta$ chains that are perpendicular to the field direction [Fig.~\ref{Fig:1}{\color{blue}(a)}]. This is a useful deconstruction as Ising pyrochlores in moderate-strength $[1,\bar{1},0]$ magnetic field possess $\alpha$ chains that are polarized to the extent allowed by the Ising single-ion anisotropy, but this field does not interact with the $\beta$-chain magnetic dipole moments which are perpendicular to the field by virtue of their single-ion Ising anisotropy. Furthermore, the symmetry of the pyrochlore lattice results in a frustrated interchain interaction between $\alpha$ and $\beta$ chains which vanishes at the nearest-neighbour level~\cite{Yoshida2004, Placke2020}, and even quantum fluctuations are not expected to result in any significant coupling between $\alpha$ and $\beta$ chains at the nearest-neighbour level~\cite{Placke2020}. 

The decoupling of $\alpha$ and $\beta$ chains in a $[1,\bar{1},0]$ magnetic field has indeed been reported for Ce$_2$Zr$_2$O$_7$~\cite{Smith2023}, Ho$_2$Ti$_2$O$_7$~\cite{Harris1997, Melko2004, Fennell2005, Clancy2009}, and Dy$_2$Ti$_2$O$_7$~\cite{Yoshida2004, Fennell2002, Melko2004, Fennell2005, Ruff2005}, where for each material, the $\alpha$ chains form a polarized, non-collinear structures with net magnetic moment along the field direction. The $\beta$ chains exhibit short-ranged ferromagnetic order of the dipole magnetic moments, as shown in Fig.~\ref{Fig:1}{\color{blue}(a)}. For each of these materials, the non-collinear ferromagnetic coupling of dipole moments in the $\beta$ chains, together with the non-collinear polarization of the $\alpha$ chains, allows for enforcement of the two-in, two-out spin ice rule in $[1,\bar{1},0]$ magnetic fields. For the classical spin ices Ho$_2$Ti$_2$O$_7$ and Dy$_2$Ti$_2$O$_7$, this ferromagnetic coupling of the dipole moments is the same ferromagnetic coupling that underlies the conventional spin ice rule~\cite{Harris1997, Ramirez1999, Bramwell2001, Fennell2009, Morris2009, Bramwell2001b, Rau2015}. For the multipolar quantum spin ices such as Ce$_2$Zr$_2$O$_7$~\cite{Changlani2022, Smith2022, Smith2023}, spin ice ground states can occur even with antiferromagnetic coupling between dipole moments~\cite{Benton2020, Patri2020, Huang2018b, Kim2022, Desrochers2022}. Nonetheless, previous estimates of the exchange parameters for Ce$_2$Zr$_2$O$_7$ indeed indicate a ferromagnetic coupling of dipole moments (Refs.~\cite{Changlani2022, Smith2022, Smith2023}) consistent with the reported magnetic structure~\cite{Smith2023}. 

For the case of Ho$_2$Ti$_2$O$_7$ and Dy$_2$Ti$_2$O$_7$, short-ranged antiferromagnetic interchain correlations between the $\beta$ chains have been reported (Refs.~\cite{Harris1997, Fennell2002, Fennell2005, Yoshida2004, Ruff2005, Clancy2009, Melko2004, Placke2020}) and have been attributed (Ref.~\cite{Yoshida2004}) to long-ranged dipolar interactions which are relatively-large due to the large magnetic moments in these materials ($\sim10$~$\mu_B$~\cite{Rosenkranz2000, Ruminy2016, Gaudet2018, Jana2002}). For the case of Ce$_2$Zr$_2$O$_7$ and its smaller magnetic moment of 1.29~$\mu_B$~\cite{Gao2019,Gaudet2019}, the long-ranged dipolar interaction, which scales with the square of the moment size, is much weaker and no interchain correlations between $\beta$ chains have been observed~\cite{Smith2023}.

Earlier neutron scattering measurements of the $\beta$-chain scattering from Ce$_2$Zr$_2$O$_7$ in Ref.~\cite{Smith2023} only have significant sensitivity to the dipole magnetic moments. Thus no detailed information exists regarding the correlations between octupolar moments associated with the $x$ and $y$ components of the pseudospin-$1/2$ in Ce$_2$Zr$_2$O$_7$, which likely correlated over longer distances than are the $z$ components, due to their stronger exchange coupling~\cite{Smith2022,Smith2023}.

For an Ising pyrochlore in a magnetic field along the $[1,1,1]$ direction, each magnetic ion in the lattice has a component of their magnetic dipole moment along the field direction. A moderate-strength magnetic field along the $[1,1,1]$ direction selects the three-in, one-out magnetic structure for each tetrahedron, as shown in Fig.~\ref{Fig:1}{\color{blue}(b)}, where each magnetic dipole moment favors the local easy-axis direction that best satisfies the Zeeman term in the Hamiltonian. The field-dependence of the magnetization of Ce$_2$Zr$_2$O$_7$ at $T=2$~K has been reported for a magnetic field along the $[1,1,1]$ direction for field strengths up to 8~T (Ref.~\cite{Gao2022}). While the maximum field-strength of 8~T was not large enough to saturate the magnetization at $T=2$~K, it is near-saturation at 8~T and approaches a value near that expected for the three-in, one-out magnetic structure~\cite{Gao2022}. This three-in, one-out phase has also been detected at low temperature for the classical spin ices Ho$_2$Ti$_2$O$_7$ (Refs.~\cite{Fennell2007, Toews2013, Krey2012}) and Dy$_2$Ti$_2$O$_7$ (Refs.~\cite{Tabata2006, Fukzawa2002, Aoki2004, Hiroi2003, Takatsu2013, Matsuhira2002, Higashinaka2004}) in moderate-strength magnetic fields along the $[1,1,1]$ direction. 

In weaker magnetic fields along the $[1,1,1]$ direction, some spin ice materials are known to show a Kagome ice phase where the spin ice rules are satisfied such that Kagome layers perpendicular to the $[1,1,1]$ direction remain frustrated, with each Kagome layer exhibiting an alternating pattern of two-in, one-out and one-in, two-out triangles~\cite{Xie2014, Kao2016, Moessner2003, Isakov2004, Bojesen2017}. This Kagome ice phase has been detected in the classical spin ices Ho$_2$Ti$_2$O$_7$ (Refs.~\cite{Fennell2007, Toews2013, Krey2012}) and Dy$_2$Ti$_2$O$_7$ (Refs.~\cite{Tabata2006, Fukzawa2002, Higashinaka2003, Aoki2004, Hiroi2003, Takatsu2013, Matsuhira2002, Tabata2007, Higashinaka2004}). Interestingly, the dipole-octupole pyrochlore Nd$_2$Zr$_2$O$_7$, which has an all-in, all-out ordered ground state in zero-field (Refs.~\cite{Lhotel2015, Xu2015, Xu2016, Opherden2017, Xu2018}) that is proximate to a quantum spin ice phase (Refs.~\cite{Petit2016,Benton2016,Xu2020,Leger2021}), also shows evidence for a Kagome ice phase in weaker magnetic fields along the $[1,1,1]$ direction~\cite{Lhotel2018}. 

The different Ce$^{3+}$-based DO pyrochlores (Ce$_2$Zr$_2$O$_7$, Ce$_2$Sn$_2$O$_7$, and Ce$_2$Hf$_2$O$_7$) have all been proposed as possessing QSI ground states, with varying degrees of agreement among the relevant studies~\cite{Smith2022, Smith2023, Changlani2022, Sibille2020, Yahne2022, Poree2023b}. Furthermore, Ce$_2$Zr$_2$O$_7$ in particular, is sensitive to oxidation and concomitant replacement of magnetic Ce$^{3+}$ with non-magnetic Ce$^{4+}$ (Ref.~\cite{Gaudet2019}), which has the potential to significantly alter the local magnetic behaviour probed in muon spin relaxation and rotation ($\mu$SR) measurements. In addition, no transverse field $\mu$SR measurements have been reported to date for any of Ce$_2$Zr$_2$O$_7$, Ce$_2$Hf$_2$O$_7$, or Ce$_2$Sn$_2$O$_7$. This motivates our present $\mu$SR study of carefully-prepared Ce$_2$Zr$_2$O$_7$ single crystal samples.

We compare our results for zero magnetic field on single crystal Ce$_2$Zr$_2$O$_7$ to those measured on powder samples of Ce$_2$Zr$_2$O$_7$ in Ref.~\cite{Gao2019} and on powder samples of Ce$_2$Sn$_2$O$_7$ in Ref.~\cite{Sibille2015}. We also compare our results for a longitudinal field to those measured on powder samples of Ce$_2$Zr$_2$O$_7$ in Ref.~\cite{Gao2019}. Our zero-field results show no signs of magnetic ordering or spin freezing, consistent with previous zero-field $\mu$SR measurements on both Ce$_2$Zr$_2$O$_7$ (Ref.~\cite{Gao2019}) and Ce$_2$Sn$_2$O$_7$ (Ref.~\cite{Sibille2015}). We measure a more gentle relaxation rate for Ce$_2$Zr$_2$O$_7$ in zero-field at low temperatures than was previously reported in Ref.~\cite{Gao2019}, and we report little temperature dependence of this small relaxation. We speculate that this difference originates from a decreased sample oxidation in our single crystal samples compared with the powder samples in Ref.~\cite{Gao2019}. In turn, we associate this with the careful annealing and storing procedures taken with our single crystal samples. The longitudinal field results show that the magnetic moments in Ce$_2$Zr$_2$O$_7$ remain dynamic on the $\mu$s time scale down to at least $T = 0.1$~K. 

For both $[1,\bar{1},0]$ and $[1,1,1]$ magnetic fields, our transverse field measurements of the temperature dependence of the $\mu$SR Knight shift in Ce$_2$Zr$_2$O$_7$ in a 1~T field deviates from the temperature dependence of the uniform susceptibility in a 0.1~T field along the $[1,\bar{1},0]$ direction below T$\sim$ 1 K.  Both of these temperature dependencies, that of the Knight shift in a 1 T magnetic field, and the uniform susceptibility in a 0.1 T magnetic field are reasonably well described quantitatively by numerical-linked-cluster (NLC) calculations using the nearest-neighbour exchange parameters estimated for Ce$_2$Zr$_2$O$_7$ in Ref.~\cite{Smith2023}. 

\section{\label{sec:Methods}Methods}

\label{sec:CZOMeth}

\subsection{\label{subsec:ExpSetup}Experimental Setup}

We present $\mu$SR measurements performed on two different high-quality single crystal samples of Ce$_2$Zr$_2$O$_7$, each grown by floating zone image furnace techniques as described in Ref.~\cite{Gaudet2019}. As also described in Ref.~\cite{Gaudet2019}, sample oxidation and accompanying non-magnetic Ce$^{4+}$ impurities can complicate the interpretation of measurements performed on as-grown Ce$_2$Zr$_2$O$_7$ samples. Furthermore, significant sample oxidation occurs for Ce$_2$Zr$_2$O$_7$ left in air at room temperature~\cite{Gaudet2019}, at least for powder samples. Accordingly, and similar to the process described in Ref.~\cite{Smith2022}, our crystals were each annealed at 1350~$^\circ$C for 72 hours in a 5\%~H$_2$, 95\%~Ar gas mixture to reduce the as-grown oxygen and Ce$^{4+}$ content. Samples were stored in inert gas after annealing (and between measurements) until being loaded into the inert environment used for measurement. These single crystal samples of Ce$_2$Zr$_2$O$_7$ 

We performed zero field (ZF), high transverse field (TF), and longitudinal field (LF) $\mu$SR measurements on the M15 and M20 beam lines at the TRIUMF Laboratory in Vancouver, Canada. On the M15 beam line, we used a spectrometer incorporating a dilution refrigerator, allowing measurements in the temperature range from 0.02~K~to~10~K. The experimental setup used a superconducting magnet to allow for fields up to 5~T. We mounted the samples on a silver cold finger using copper grease mixed with Apiezon N grease to ensure adequate heat transfer between the sample and the cold finger. Silver gives a well-understood $\mu$SR background signal. The spectrometer on the M15 beam line has a time resolution of 0.4~ns. On the M20 beam line, we used the low background apparatus, LAMPF, with temperature capabilities from 1.9~K~to~300~K and fields up to 0.4~T. The LAMPF instrument also has a resolution of 0.4~ns. We took two sets of measurements with different samples for each: One set with the initial muon polarization along the $[1,\bar{1},0]$ direction and a second set with the initial muon polarization along the $[1,1,1]$ direction. Each large single crystal sample was aligned using an x-ray Laue diffractometer and was cut into several 2~mm thick disks with the face of each disk perpendicular to the desired direction of muon polarization: One set of disks with faces perpendicular to the $[1,\bar{1},0]$ direction and a second set of disks with faces perpendicular to the $[1,1,1]$ direction. The resulting single crystal disks of Ce$_2$Zr$_2$O$_7$ were transparent, consistent with the high crystallinity of the sample that was also determined through our x-ray Laue measurements. The aforementioned annealing and sealing procedure (see previous paragraph) was performed on these single crystal disks of Ce$_2$Zr$_2$O$_7$ and importantly, the annealed samples had a yellow-green colour, consistent with a high Ce$^{3+}$ content and low Ce$^{4+}$ content.

\subsection{\label{subsec:DataAnalysis}Data Analysis}

We used the muSRfit software platform to analyse the collected $\mu$SR spectra ~\cite{Suter2012}. A weak transverse field ($\sim0.002$~T) measurement was used to measure the total muon asymmetry, $A_{0}$, as well as the detector- and sample-position dependent correction, $\alpha$. On the M15 beam line, TF measurements were used to estimate the fraction of the signal coming from the sample, $F_{\mathrm{s}}$, versus the silver cold finger as the two signals are readily discernible using the muon Knight shift of the TF measurements. The values we measured for $F_{\mathrm{s}}$ are 0.69 for the $[1,\bar{1},0]$-field data and 0.70 for $[1,1,1]$-field data, which are consistent with other samples prepared in the same manner where the sample signal fraction is 0.6 - 0.75 in ZF. As the measured relaxation rate of Ce$_{2}$Zr$_{2}$O$_{7}$ is comparable to silver, and the asymmetry is relatively featureless, we cannot use our ZF data to determine the fraction of the signal coming from our sample. Instead, we use the TF data to give our best estimate of the sample fraction in ZF. The silver relaxation rate used in our ZF analysis was fit from previous measurements performed on the silver sample holder alone. 

To directly compare our ZF results to the results from Gao \emph{et al}. in Ref.~\cite{Gao2019}, the asymmetry signals of both data sets were converted to the muon polarization. The advantage of displaying the muon polarisation is that instrumental differences between the two measurements are removed to a high degree. For each data set, this calculation requires the total asymmetry, $A_{0}$, and the fraction of the asymmetry coming from the sample, $F_{\mathrm{s}}$, to be known. We use the values listed in Ref.~\cite{Gao2019} and those from our own measurements to determine these. The muon polarization can then be expressed as
\begin{equation}\label{eq:P}
    P_{\mu}\left(t\right) = \frac{A\left(t\right)-\left(1-F_{\mathrm{s}}\right)A_{0}}{A_{0} F_{\mathrm{s}}},
\end{equation}
where $A(t)$ is the measured asymmetry signal. 

To calculate the muon Knight shift, $K$, in our TF measurements, we use the field measured by muons stopping in the silver sample holder as a field reference. Silver has a small, positive frequency shift of 94 ppm, roughly 20 times smaller than the smallest frequency shift measured in our samples. We then calculate the frequency shift as,

\begin{equation} \label{eq:K}
    K = \frac{B_{\mathrm{sample}}-B_{\mathrm{Ag}}}{B_{\mathrm{Ag}}},
\end{equation}
where $B_{\mathrm{sample}}$ is the magnetic field measured by muons stopping in the sample and $B_{\mathrm{Ag}}$ is the magnetic field measured by muons stopping in the sample holder.

We compare our transverse field measurements of the $\mu$SR Knight shift, which is a measure of the local value of $M/h$~\cite{Yaouanc2011}, with fourth-order NLC calculations for the magnetic susceptibility calculated via $M/h$, where $M$ is the magnetization. We also compare the $\mu$SR Knight shift (measured in a $h = 1$~T field), and our NLC calculations, with the direct current (DC) magnetic susceptibility determined according to $M/h$ using the measured bulk magnetization in a $h = 0.1$~T field.

Methods based on exact diagonalization and matrix product states are limited due to the three-dimensional nature, and quantum Monte-Carlo faces the infamous sign problem hindering its application. However, at a finite temperature, the NLC method has proven valuable in studying frustration in three dimensions~\cite{Yahne2022, Smith2022, Smith2023}. The NLC method uses the exact diagonalization of clusters, with truncation at some maximum cluster size, in order to calculate extensive quantities in a spirit similar to a finite temperature expansion. The order of a specific NLC calculation is defined as the maximum number of tetrahedra considered in a cluster, and for a given order $n$, the calculation is near exact when it converges with the corresponding $(n-1)$th order calculation. We use the region of convergence to define a low temperature threshold for each calculation, for which the calculation has high accuracy above the threshold and is not shown below the threshold where the accuracy is unknown. Examples of the NLC method applied to the dipole-octupolar pyrochlores Ce$_{2}$Zr$_{2}$O$_{7}$ and Ce$_{2}$Sn$_{2}$O$_{7}$ can be found in Refs.~\cite{Yahne2022, Smith2022, Smith2023}, and further details on the NLC method are provided in Ref.~\cite{Applegate2012, Tang2013, Tang2015, Schafer2020, RobinThesis}. We use fourth-order NLC calculations in order to calculate the magnetic susceptibility for both $[1,\bar{1},0]$ and $[1,1,1]$ magnetic fields of strength $h=0.1$~T and $h=1$~T, and we use Euler transformations to improve convergence (see Ref.~\cite{Applegate2012, Smith2022} for example) for each calculation.

\section{\label{sec:Results}Results and Discussion}

\subsection{\label{sec:ZF_Results}Zero Field Measurements}

\begin{figure}[t]
\linespread{1}
\par
\includegraphics[width=3.4in]{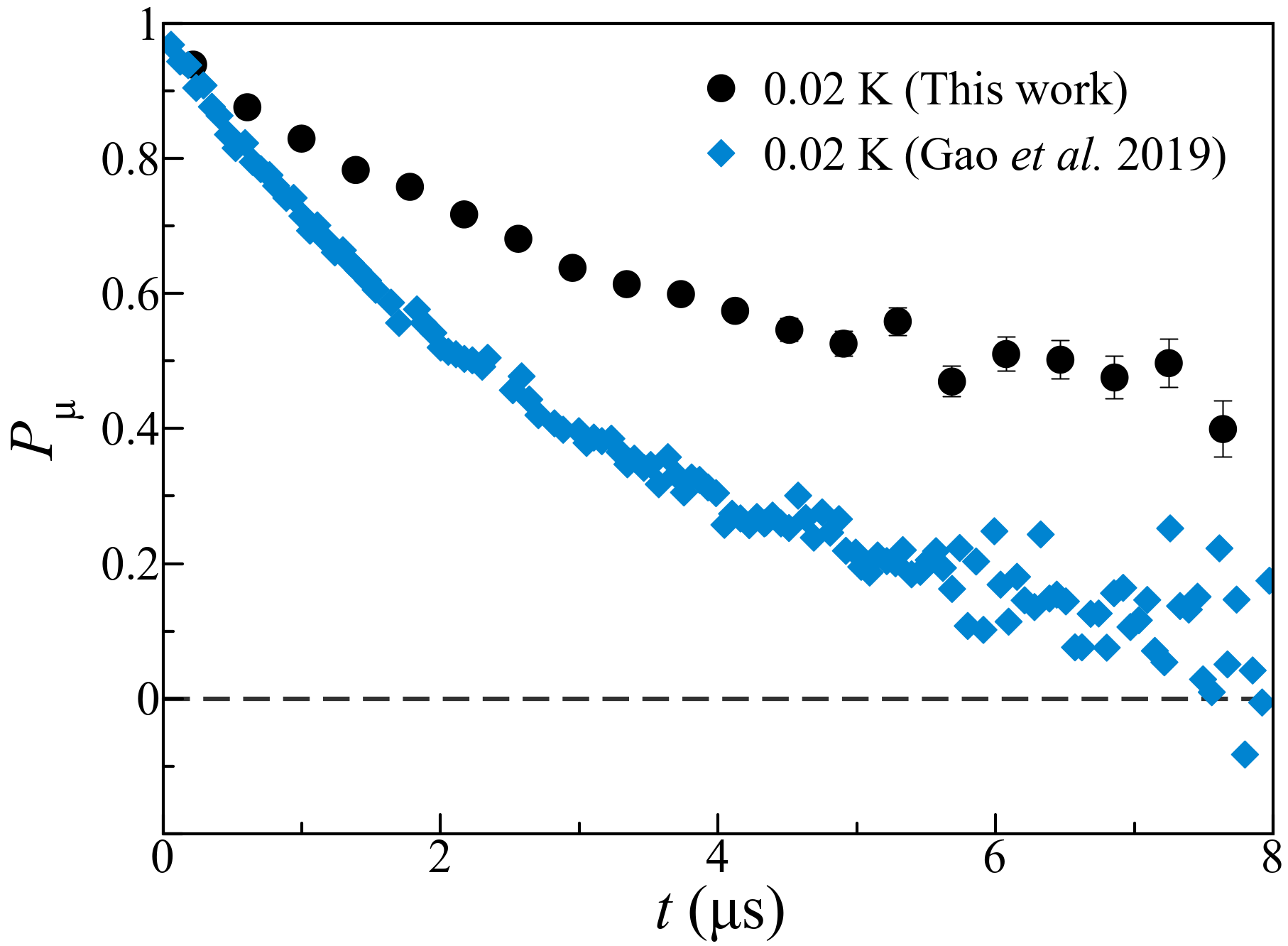}
\par
\caption{A comparison of the muon polarization in zero-field and at low temperature ($T = 0.02$~K) measured from single crystal Ce$_{2}$Zr$_{2}$O$_{7}$ in this work (black), and from powder Ce$_{2}$Zr$_{2}$O$_{7}$ in Ref.~\cite{Gao2019} (blue). Stronger relaxation is evident in the ZF $\mu$SR data from the powder sample.} 
\label{Fig:2}
\end{figure}

We performed zero-field $\mu$SR measurements to search for the presence of static magnetic moments in our single crystal samples of Ce$_{2}$Zr$_{2}$O$_{7}$. Measurements on our single crystal samples were performed with the initial muon polarization along the $[1,1,1]$ direction. As shown in Figs.~\ref{Fig:2} and ~\ref{Fig:3}, no sign of magnetic ordering or spin freezing was observed to temperatures as low as $T = 0.02$~K, as would be evidenced from either oscillations in the muon polarization at low temperature [Fig.~\ref{Fig:2}], or from a sudden and large increase in the relaxation rate as a function of temperature [Fig.~\ref{Fig:3}{\color{blue}(a,c)}].

While ZF measurements on powder samples of Ce$_{2}$Zr$_{2}$O$_{7}$ by Gao \emph{et al} (Ref.~\cite{Gao2019}) similarly show no oscillations at low temperatures, as also displayed in Fig.~\ref{Fig:2}, they do observe a significant increase in the relaxation rate below $\sim$0.5~K,  which is not observed in our single crystal sample [Fig.~\ref{Fig:3}{\color{blue}(c)}].  Our ZF measurements show little change at all from 4~K to 0.02~K, as shown in Fig.~\ref{Fig:3}{\color{blue}(a)}, and in particular, little change in the relaxation rate as a function of temperature at all temperatures measured, as shown in Fig.~\ref{Fig:3}{\color{blue}(c)}.  

Equation~\ref{eq:P} is used to convert the measured asymmetry data to muon polarization as described in Section~\ref{subsec:DataAnalysis}. The much reduced relaxation rate at base temperature ($T = 0.02$~K) in our new ZF measurements from single crystals is evident in Fig.~\ref{Fig:2}, as compared to comparable measurements on powder samples by Gao \emph{et al.} in Ref.~\cite{Gao2019}. These two sets of muon polarization data are plotted on the same scale in Fig.~\ref{Fig:2} for ease of comparison.  The more gentle ZF relaxation observed in the data from our annealed, single crystal samples of Ce$_{2}$Zr$_{2}$O$_{7}$, implies that there is an extra source of relaxation in the Ref.~\cite{Gao2019} powder samples that is not present in our single crystal samples. We speculate that this is likely due to a higher degree of oxidation in the powder samples as compared with our single crystal sample, which originates from the much smaller surface-to-volume ratio of the single crystal compared with the powder sample, and the careful sample preparation undertaken to ensure the cerium ions are in the 3+ oxidation state in our single crystal samples. This sample preparation included annealing our samples in hydrogen and storing them in inert gas before each experiment as described in Section~\ref{subsec:ExpSetup}. Powder samples of Ce$_{2}$Zr$_{2}$O$_{7}$ have been shown to start oxidizing after 1~hour in air at room temperature~\cite{Gaudet2019}. 

Our ZF results [Figs.~\ref{Fig:2}, \ref{Fig:3}{\color{blue}(a-c)}] are more reminiscent of $\mu$SR measurements performed by Sibille \emph{et al.} on another dipole-octupole pyrochlore, Ce$_{2}$Sn$_{2}$O$_{7}$~\cite{Sibille2015, Sibille2020}, where the muon spin polarization as a function of time remains largely unchanged over the range of measured temperatures. Our ZF $\mu$SR measurement results are fit to the same phenomenological, stretched exponential model as in Refs.~\cite{Gao2019, Sibille2015}, 
\begin{equation}\label{eq:CZO_ZF}
    A(t)/ A_{0} = F_{\mathrm{s}} e^{-\left( \lambda t\right)^{\beta}} + \left(1-F_{\mathrm{s}}\right)e^{- \lambda_{\mathrm{Ag, ZF}} t},
\end{equation}
where, $A_{0}$ is the total asymmetry, $F_{\mathrm{s}}$ is the fraction of the signal coming from the sample, $\lambda$ is the temperature-dependent relaxation rate of the sample, $\beta$ is the temperature-dependent stretching exponent, and $\lambda_{\mathrm{Ag, ZF}}$ is the relaxation rate of silver in zero field. Figure~3{\color{blue}(a)} shows our ZF asymmetry data fit to Eq.~\ref{eq:CZO_ZF}, where a weak temperature dependence is observed. The black points in Fig.~\ref{Fig:3}{\color{blue}(c)} show the extracted relaxation rate in our single crystal of Ce$_{2}$Zr$_{2}$O$_{7}$, and the fact that it changes very little over our measured temperature range. This can be compared to Ref.~\cite{Gao2019} where the relaxation rate is larger and shows a stronger temperature dependence, with the relaxation rate increasing sharply at temperatures below $\sim$ 0.5 K. Our measured stretching exponent is shown in Fig.~\ref{Fig:3}{\color{blue}(b)} and it remains relatively constant below 2~K between 0.63 and 0.68 while above 2~K the exponent is closer to the expected value for a dilute paramagnet ($\beta = 1$~\cite{Yaouanc2011}). In contrast, the stretching exponent as measured in Ref.~\cite{Gao2019} shows a sudden increase at the lowest temperatures. 

Our results on single crystal Ce$_{2}$Zr$_{2}$O$_{7}$ more closely resemble those on powder Ce$_{2}$Sn$_{2}$O$_{7}$~\cite{Sibille2015}, where the relaxation rate [Fig.~\ref{Fig:3}{\color{blue}(c)}] is constant at $\lambda \approx$ 0.05~$\mu s^{-1}$ and the stretching exponent [Fig.~\ref{Fig:3}{\color{blue}(b)}] is constant at $\beta\approx$ 0.5 below 0.8 K. Studies of Ce$_{2}$Sn$_{2}$O$_{7}$ have shown that this material is stable in air and therefore more robust to Ce$^{4+}$ defects as compared to Ce$_{2}$Zr$_{2}$O$_{7}$ powder~\cite{Sibille2020, Yahne2022}. 

The increased relaxation at low temperature in the Ce$_{2}$Zr$_{2}$O$_{7}$ powder samples of Ref.~\cite{Gao2019}, compared to the single crystals in this work, suggests that samples from Ref.~\cite{Gao2019} possess enhanced slow (low energy) spin fluctuations compared to the annealed single crystal that is the subject of the present work. Furthermore, the different sample-preparation methods used in this work (see Sec.~\ref{subsec:ExpSetup}) and in Ref.~\cite{Gao2019} suggests that this difference in the low-temperature spin fluctuations is due to additional impurities associated with sample oxidation in the samples of Ref.~\cite{Gao2019}, compared to samples which are prepared through the annealing and sealing procedures used in this work~\cite{Gaudet2019,Smith2022,Smith2023}.


It is worth mentioning that the magnetic field due to an octupole moment falls off with distance as $\frac{1}{r^{5}}$, significantly faster than that of a dipole moment, which falls off as $\frac{1}{r^{3}}$~\cite{Griffiths1999}. Therefore the $\mu$SR sensitivity to octupolar moments should be much weaker than that to dipole moments, and we do not expect the present measurements to have significant sensitivity to the octupolar degrees of freedom.

\subsection{\label{sec:LF_Results}Longitudinal Field Measurements}


\begin{figure*}[t]
\linespread{1}
\par
\includegraphics[width=6in]{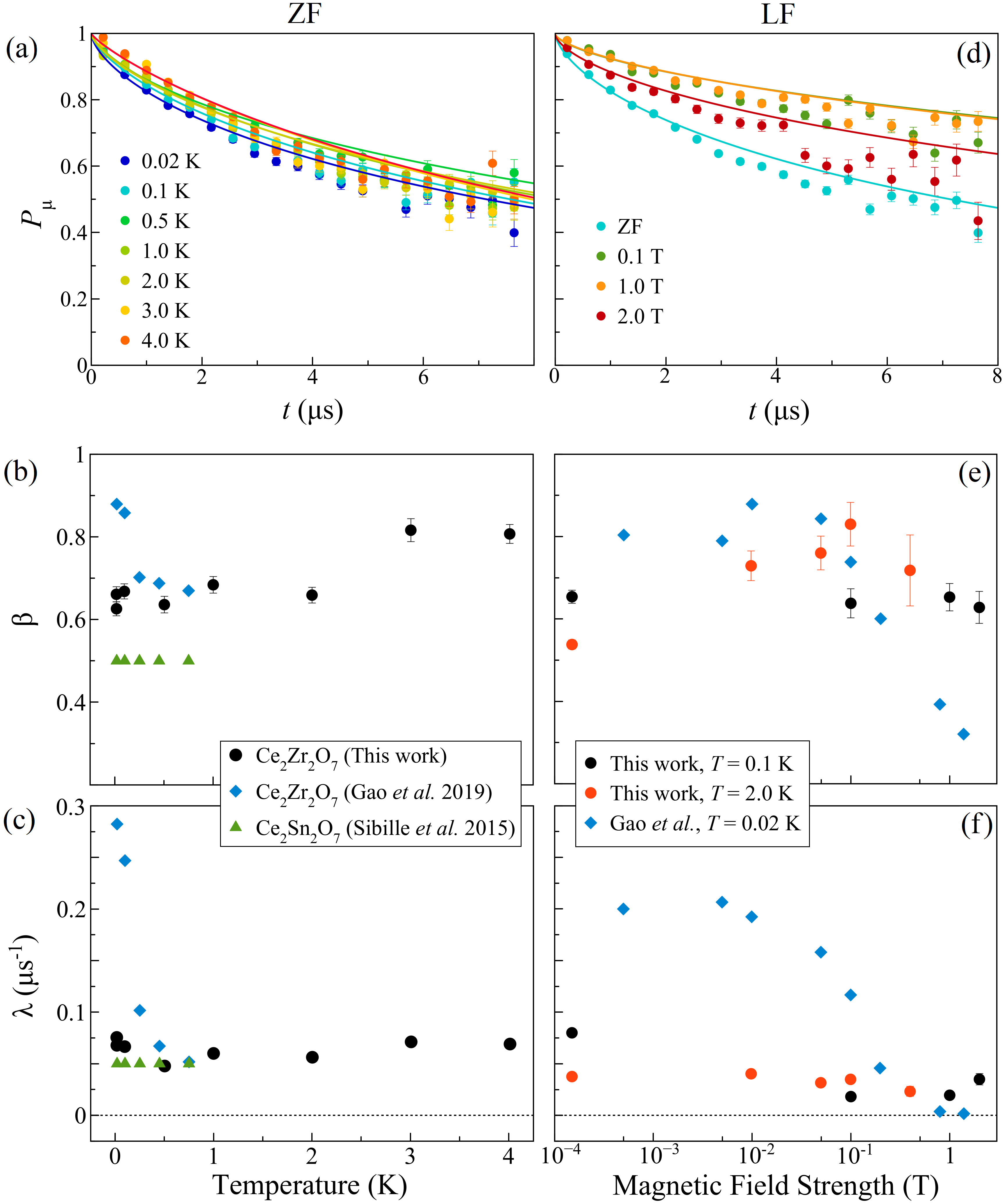}
\par
\caption{(a)~The muon polarization measured from our single crystal Ce$_{2}$Zr$_{2}$O$_{7}$ samples in zero-field, fit to Eq.~\ref{eq:CZO_ZF}. (b)~The stretching exponent and (c) the relaxation rate measured from our single crystal Ce$_{2}$Zr$_{2}$O$_{7}$ samples in zero-field (black), as well as those measured from powder Ce$_{2}$Zr$_{2}$O$_{7}$ (blue) and powder Ce$_{2}$Sn$_{2}$O$_{7}$ (green) in Refs.~\cite{Gao2019, Sibille2015}. (d)~The muon polarization measured from our Ce$_{2}$Zr$_{2}$O$_{7}$ samples at $T = 0.1$~K in a magnetic field along the $[1,1,1]$ direction with the muon polarization along the field, fit to Eq.~\ref{eq:CZO_ZF} with $\lambda_{\mathrm{Ag, ZF}}$ set to zero for non-zero-field. (e)~The longitudinal field stretching exponent and (f) the longitudinal field relaxation rate measured from our single crystal Ce$_{2}$Zr$_{2}$O$_{7}$ samples at $T = 0.1$~K on the M15 beam line (black) and at $T = 2$~K on the M20 beam line (red), as well as those measured from powder Ce$_{2}$Zr$_{2}$O$_{7}$ at $T = 0.02$~K in Ref.~\cite{Gao2019} (blue).} 
\label{Fig:3}
\end{figure*}


To confirm the dynamic nature of the low-temperature magnetism in our samples, we performed longitudinal field measurements at $T=0.1$~K with the results shown in Fig.~\ref{Fig:3}{\color{blue}(d-f)}. Fig.~\ref{Fig:3}{\color{blue}(d)} shows that the polarization function changes only a small amount as a function of the applied field and is not fully decoupled, even in a field of 2~T along the [1,1,1] direction. This demonstrates that the small magnetic dipole moment of the Ce$^{3+}$ ions must be dynamic on the $\mu$s time scale and, in combination with the ZF result, rules out any transition to long-ranged order, or a frozen state, of the dipole moments. 

We fit our LF results to Eq.~\ref{eq:CZO_ZF} where we fix $\lambda_{\mathrm{Ag, ZF}}$ = 0 at non-zero field-strengths, as the small nuclear fields of silver will be fully decoupled above 0.002~T. The relaxation rate [Fig.~\ref{Fig:3}{\color{blue}(f)}] and stretching exponent [Fig.~\ref{Fig:3}{\color{blue}(e)}] so determined are roughly field-independent with the results of measurements on the M15 beam line taken at $T = 0.1$~K shown in black and measurements on the M20 beam line taken at $T = 2$~K shown in red. In contrast to our measurements, the results of Ref.~\cite{Gao2019} at $T = 0.02$~K show significant field dependence, with both the relaxation rate and stretching exponent decreasing with increasing field. Ref.~\cite{Sibille2015} on Ce$_{2}$Sn$_{2}$O$_{7}$ shows only one LF asymmetry distribution, which was taken in a 2~T field, and this LF distribution shows little change from the ZF distribution, similar to our data. 

The increased field-dependence of the low-temperature relaxation rate for the Ce$_{2}$Zr$_{2}$O$_{7}$ powder samples of Ref.~\cite{Gao2019} compared to the single crystals in this work suggests that samples from Ref.~\cite{Gao2019} have low-temperature spin fluctuations at a lower energy than the low-temperature spin fluctuations in the samples of the present work. This is consistent with conclusions drawn from the ZF measurements in Sec.~\ref{sec:ZF_Results} and the difference in sample-preparation methods used in this work and in Ref.~\cite{Gao2019} again suggests that increased oxidation-related impurities play a role in lowering the energy of spin fluctuations in the samples of Ref.~\cite{Gao2019} compared to the samples of the present work.

Our fits to the asymmetry, both for zero-field in Fig.~\ref{Fig:3}{\color{blue}(a)} and for longitudinal field in Fig.~\ref{Fig:3}{\color{blue}(d)}, fit better at early times due to the smaller error bars at these times and the inclusion of the error bars in our goodness-of-fit parameter. To investigate the robustness of our results, we repeated the same analysis shown in Fig.~\ref{Fig:3} but excluding the error bars in our least-squares goodness-of-fit parameter. This alternative fitting provides more visually-appealing fits to the ZF and LF asymmetry with the data fit equally well at low and high relaxation time.  It also produced fit parameters which were similar to those obtained with the proper inclusion of the error bars in the fits.

It is not so surprising that a stretched exponential function, which is a phenomenological model, does not completely describe our data. However, its use is standard in the field when the spin polarization function is relatively featureless, as ours is. Due to the absence of a Kubo-Toyabe-like dip in our data, more general, theoretically motivated functions such as a dynamic Lorentzian or dynamic Gaussian Kubo-Toyabe (and related functions) have a large parameter space that adequately fits the data, making it difficult to extract meaningful information from such fits. In the present case of Ce$_{2}$Zr$_{2}$O$_{7}$, these more general functions did not provide a significantly better description of the data compared with Eq.~\ref{eq:CZO_ZF}.

\subsection{\label{sec:TF_Results}Transverse Field Measurements}


\begin{figure*}[t]
\linespread{1}
\par
\includegraphics[width=6in]{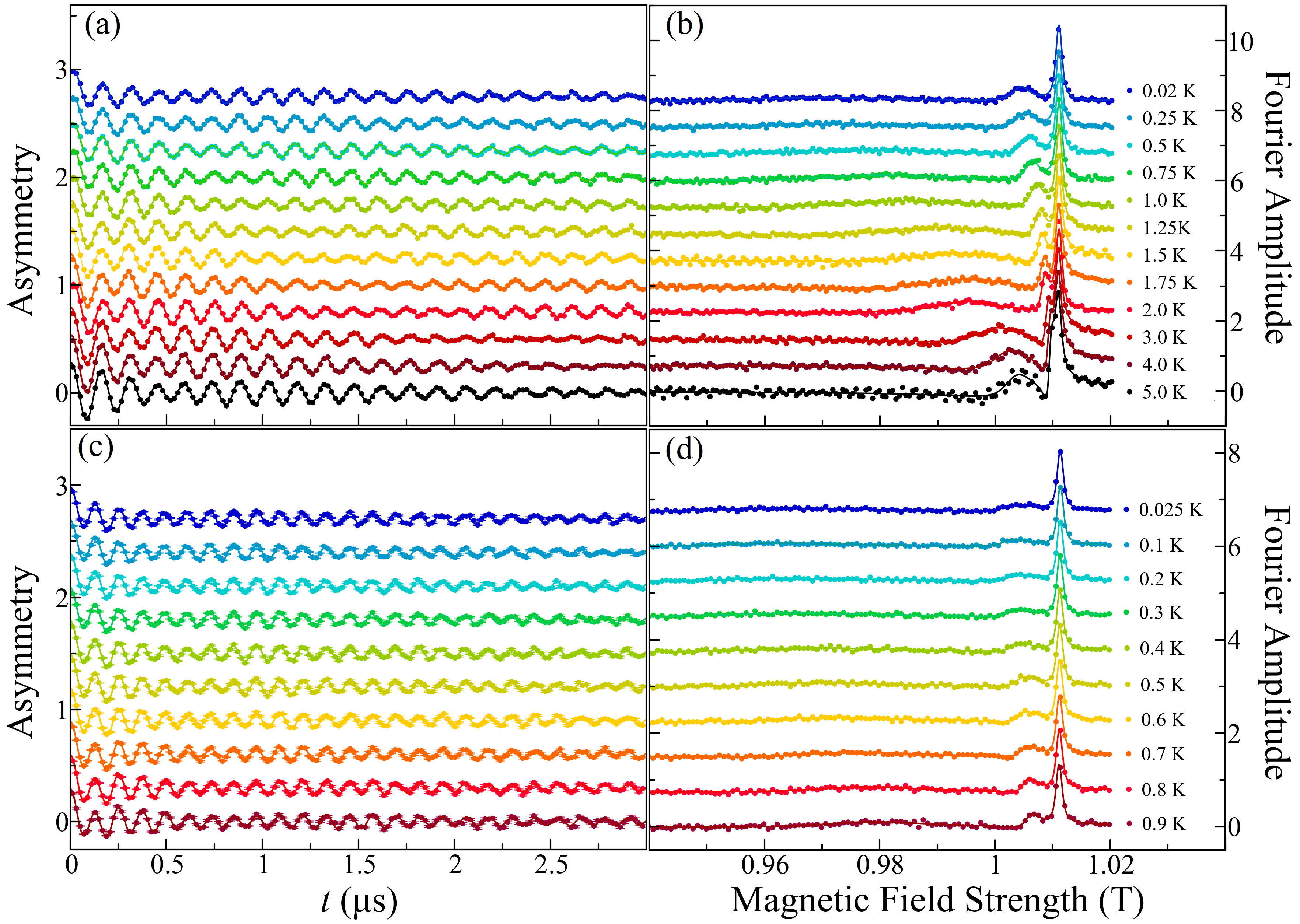}
\par

\caption{The asymmetry, $A(t)$, measured from our single crystal Ce$_{2}$Zr$_{2}$O$_{7}$ samples at various temperatures between 0.02~K and 5~K (as labeled) in a 1~T magnetic field transverse to the muon polarization, for fields along the (a) $[1,1,1]$ and (c) $[1,\bar{1},0]$ directions and for a rotating reference frame of 0.95~T. We also show the Fourier transform of the asymmetry of Ce$_{2}$Zr$_{2}$O$_{7}$ in a 1~T field along the (b) $[1,1,1]$ and (d) $[1,\bar{1},0]$ directions to demonstrate the different components of the signal for illustrative purposes. Each data set, except the highest-temperature data set of each plot, has been shifted vertically for aesthetic purposes.} 
\label{Fig:4}
\end{figure*}


Figs.~\ref{Fig:4}{\color{blue}(a,c)} shows our Ce$_{2}$Zr$_{2}$O$_{7}$ high transverse field (TF) precession data in a rotating reference frame of 0.95~T fit to the sum of three exponentially decaying cosine functions: two corresponding to two inequivalent muon stopping sites within our sample and one corresponding to the silver sample holder. Explicitly, our TF data is fit to the equation,
\begin{equation}
    \begin{split}
   A(t) /A_{0} &= F_{\mathrm{s}}F_{1}\cos \left(\gamma_{\mu} B_{1} t \right) e^{-\lambda_{1}t} \\
   &+ F_{\mathrm{s}}\left(1-F_{1}\right)\cos \left(\gamma_{\mu} B_{2} t \right) e^{-\lambda_{2} t} \\ 
   &+\left(1-F_{\mathrm{s}}\right)\cos \left(\gamma_{\mu} B_{\mathrm{Ag}} t \right) e^{-\lambda_{\mathrm{Ag, TF}}t},   
   \end{split}
\end{equation}    
where $A_{0}$ is the initial muon asymmetry, $F_{\mathrm{s}}$ is the fraction of the signal coming from the sample, $F_{1}$ is the fraction of the signal from the sample that is coming from the first muon stopping site, $B_{1}$ ($B_{2}$) is the local field at the first (second) muon stopping site, $\lambda_{1}$ ($\lambda_{2}$) is the relaxation rate at the first (second) muon stopping site, $\gamma_{\mu}$ is the muon gyromagnetic ratio, $B_{\mathrm{Ag}}$ is the magnetic field felt by muons landing in the silver sample holder, and $\lambda_{\mathrm{Ag, TF}}$ is the relaxation rate of muons landing in the silver sample holder. In our transverse field data, we were able to fit the silver relaxation directly due to the difference in the Knight shift between Ce$_{2}$Zr$_{2}$O$_{7}$ and silver.

\begin{figure}[t]
\linespread{1}
\par
\includegraphics[width=3.4in]{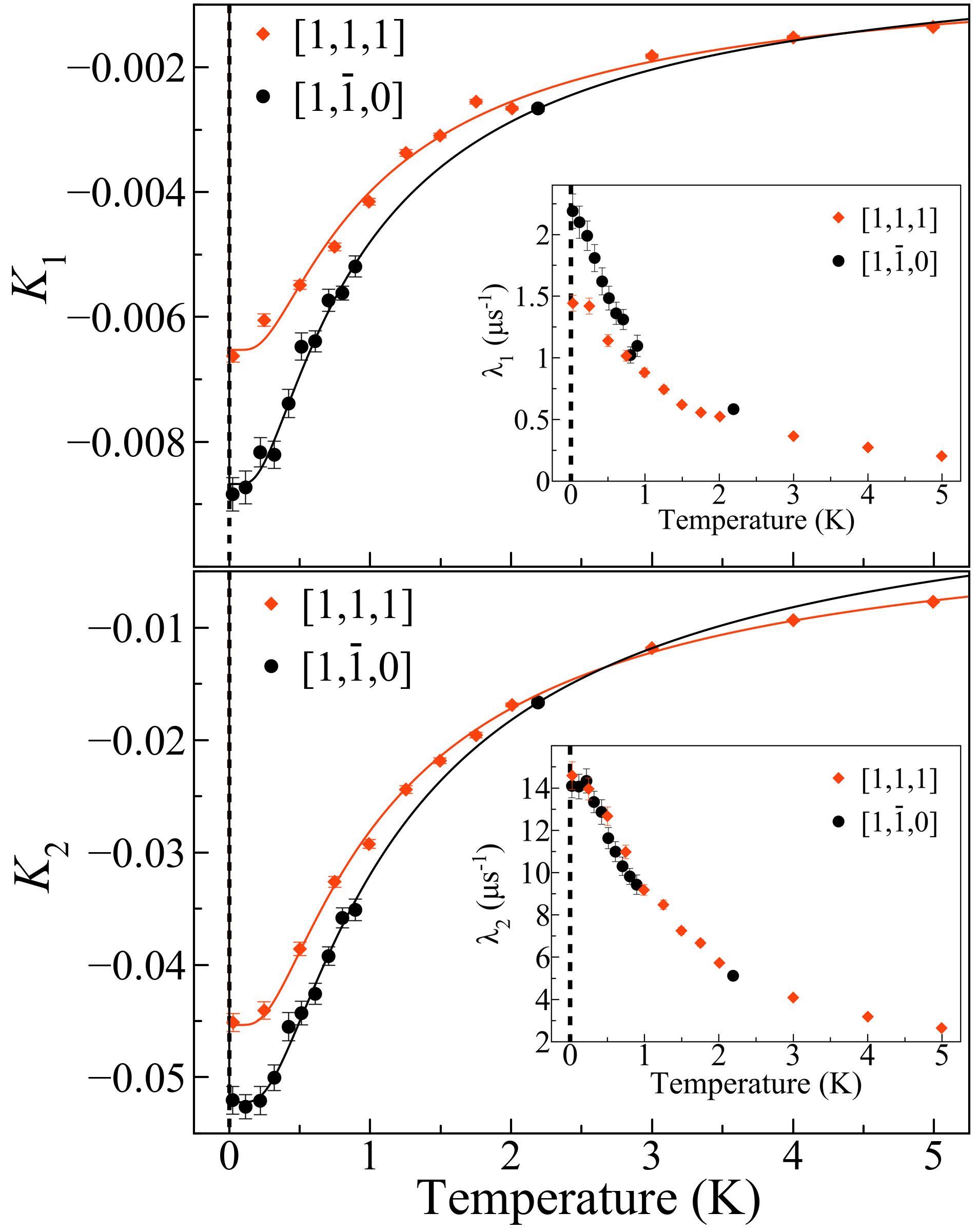}
\par
\caption{The frequency shifts and relaxation rates measured from our single crystal Ce$_{2}$Zr$_{2}$O$_{7}$ samples in a magnetic field of strength $h=1$~T along the $[1,1,1]$ (red) and $[1,\bar{1},0]$ (black) directions, transverse to the muon polarization in each case. The temperature dependence of the frequency shift of the (a) first and (b) second components of asymmetry for a transverse magnetic field along the $[1,1,1]$ and $[1,\bar{1},0]$ directions. The insets to (a) and (b) show the temperature-dependent relaxation rates of the first and second components of the asymmetry, respectively, for both field directions used in this work. The lines in (a) and (b) show our fits of the data using $K = K_0 + Ce^{-\Delta/T}$, with the best fit values of $K_0$, $C$, and $\Delta$ shown in Table~1.} 
\label{Fig:5}
\end{figure}

The three separate signals are visible in the Fourier transform of the data [Fig.~\ref{Fig:4}{\color{blue}(b,d)}] where a temperature-independent line is observed at the measured applied field of 1.01~T, corresponding to the silver sample holder, and two broader lines are observed whose centers shift to lower values and whose line-width increases as the temperature is lowered. Fits are performed in the time domain and the Fourier transform is shown for demonstrative purposes only. The resulting Knight shifts and relaxation rates are shown in Fig.~\ref{Fig:5}, where it is seen that the frequency shift (Eq.~\ref{eq:K}) for one component of the signal is about 8~times the frequency shift of another, while the relaxation is greater by a factor of about 7. The frequency shift of each component [Fig.~\ref{Fig:5}{\color{blue}(a,b)}] behaves similarly above 0.25~K with both increasing in magnitude as the temperature is lowered. Below this temperature, the frequency shift of the second component plateaus. The relaxation rates of both components [insets to Fig.~\ref{Fig:5}{\color{blue}(a,b)}] increase as the temperature is lowered to 0.25~K where the second component shows evidence of a plateau below this temperature, especially for the $[1,\bar{1},0]$ magnetic field direction where the data shows a clear and full levelling off below 0.25~K. The relaxation rate and frequency shift scale linearly with respect to one another as a function of temperature for both components, indicating that the increase in both of these values with decreasing temperature is due to the same source (see Appendix~A). 

The lines in Fig.~\ref{Fig:5} show fits to the measured Knight shifts of the form $K = K_0 + Ce^{-\Delta/T}$ suggested in Ref.~\cite{Chen2023}, where the best-fitting values of $K_0$, $C$, and $\Delta$ for each field direction measured and each component of the signal are shown in Table~1. While the form $K = K_0 + Ce^{-\Delta/T}$ has clear physical meaning in low field strengths when the zero-field QSI ground state persists (see Ref.~\cite{Chen2023}), the physical interpretation is not clear at the field strength of $h = 1$~T used for our measurements and so we employ this equation from Ref.~\cite{Chen2023} as a phenomenological form only, but it is one that describes the trend in the measured data remarkably-well. Interestingly, the best-fitting results show the relation $K_0 \approx -C$ for each field direction and component of the signal. We also note that the best-fit value of $K_0$ is slightly lower in absolute value for the $[1,\bar{1},0]$ field direction compared to the $[1,1,1]$ field direction. The best-fit value of $\Delta$ varies slightly between data sets with no systematic dependence shown between the $[1,1,1]$ and $[1,\bar{1},0]$ field direction and an average value of $\Delta \sim 0.93$~K.

It is not uncommon to observe two or more distinct inequivalent muon stopping sites in transverse field measurements on single crystal pyrochlore materials~\cite{D'Ortenzio2013, Beare2021}, as is reported here on our single crystal Ce$_2$Zr$_2$O$_7$. This may occur due to multiple (meta-)stable muon sites which are distinct in zero-field. This may also occur due to a field-induced distinction of muon sites which are equivalent in zero external field, due to changes in the magnetic structure of these pyrochlore materials as we now discuss for the case of Ce$_2$Zr$_2$O$_7$ in a $[1,\bar{1},0]$ magnetic field. 

For a magnetic field along the $[1,\bar{1},0]$ direction, it is known that the magnetic ions in the spin ice materials Dy$_2$Ti$_2$O$_7$ (Refs.~\cite{Tabata2006, Fukzawa2002, Higashinaka2003, Aoki2004, Hiroi2003, Takatsu2013, Matsuhira2002, Tabata2007, Higashinaka2004}), Ho$_2$Ti$_2$O$_7$ (Refs.~\cite{Harris1997, Melko2004, Fennell2005, Clancy2009}), and Ce$_2$Zr$_2$O$_7$ (Ref.~\cite{Smith2023}) can be decomposed into independent chains of spins, referred to as $\alpha$ and $\beta$ chains, whose behavior in this field is very different from one another, as illustrated in Fig.~\ref{Fig:1}{\color{blue}(a)} and described in Sec.~\ref{sec:Intro}. The TF $\mu$SR Knight shifts and relaxation rates are only sensitive to the component of the magnetic field along the applied field direction, which is along the $[1,\bar{1},0]$ direction in this case. Since the net dipole moment of the $\beta$ chains is perpendicular to this direction (along the $[1,1,0]$ direction)~\cite{Smith2023}, the $\beta$ chains likely create much smaller magnetic fields along the $[1,\bar{1},0]$ direction at potential muon stopping sites compared to that produced by $\alpha$ chains which exhibit non-collinear polarization along the $[1,\bar{1},0]$ direction. Similarly, the $\beta$ chains, which do not couple to magnetic fields at first approximation, should have a much smaller susceptibility, and therefore, a smaller Knight shift, compared to the $\alpha$ chains. Accordingly, we attribute component 1 and component 2 of the measured signal to muon stopping sites 1 and 2, respectively, and we speculate that muon stopping site 1 is significantly further away from the nearest $\alpha$ chain than stopping site 2 is, where both of these sites give the same signal in zero field (when there is no distinction between the $\alpha$ and $\beta$ chains). A similar attribution of muon sites may be possible for the three-in, one-out structure [Fig.~\ref{Fig:1}{\color{blue}(b)}] expected for moderate-strength $[1,1,1]$ magnetic fields, such that, again the distinction between the two sites goes away in zero field.

\begin{figure}[t]
\linespread{1}
\par
\includegraphics[width=3.4in]{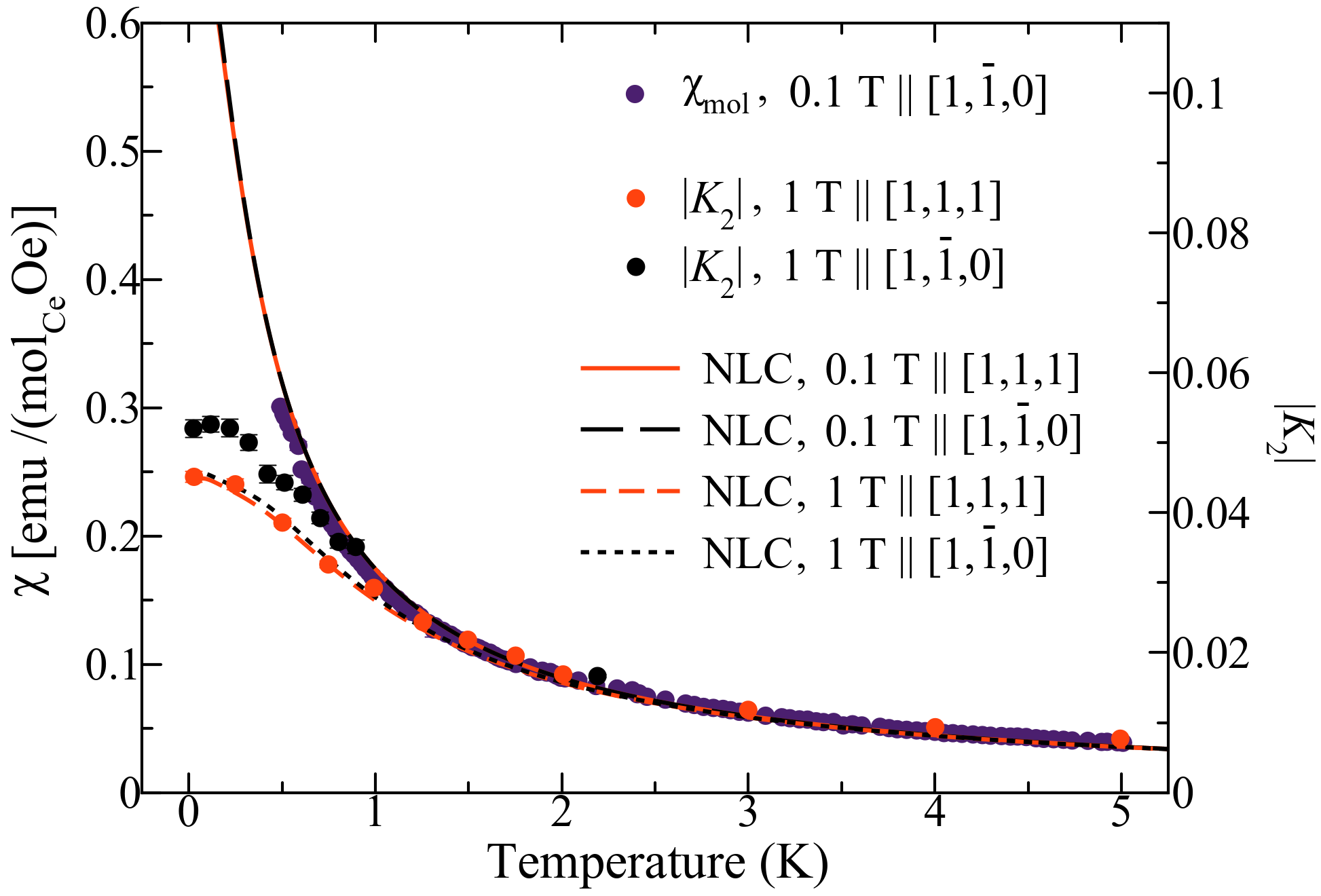}
\par
\caption{A comparison of the DC magnetic susceptibility measured from one of our single crystal samples of Ce$_{2}$Zr$_{2}$O$_{7}$ for a field of strength $h=0.1$~T along the $[1,\Bar{1},0]$ direction (purple, left axis), with the absolute value of the Knight shift corresponding to the second component of asymmetry measured from our single crystal samples of Ce$_{2}$Zr$_{2}$O$_{7}$ for a transverse magnetic field of strength $h=1$~T along the $[1,1,1]$ (red, right axis) and $[1,\bar{1},0]$ (black, right axis) directions. The lines in this plot show our fourth-order NLC calculations of $M/h$ for both a $[1,1,1]$ (red, left axis) and $[1,\bar{1},0]$ (black, left axis) magnetic field, for field strengths of $h=0.1$~T and $h=1$~T (as labelled), calculated using the nearest-neighbour exchange parameters estimated for Ce$_{2}$Zr$_{2}$O$_{7}$ in Refs.~\cite{Smith2022, Smith2023}, $(J_{\tilde{x}}, J_{\tilde{y}}, J_{\tilde{z}}) = (0.063, 0.062, 0.011)$~meV and $\theta = 0$, with $g_z$ fit to $g_z = 2.38$ using only the $h=0.1$~T susceptibility data.}
\label{Fig:6}
\end{figure}

Figure~6 shows a comparison of the $\mu$SR Knight shift measured from Ce$_{2}$Zr$_{2}$O$_{7}$ at $h = 1$~T, with the bulk susceptibility measured from Ce$_{2}$Zr$_{2}$O$_{7}$ at $h = 0.1$~T. Specifically, Fig.~\ref{Fig:6} shows the bulk susceptibility, obtained from DC magnetic susceptibility measurements on one of our single crystal samples of Ce$_{2}$Zr$_{2}$O$_{7}$ with a field of strength $h=0.1$~T along the $[1,\Bar{1},0]$ axis, and the magnetic susceptibility measured as the absolute value of the Knight shift corresponding to the second component of asymmetry measured from Ce$_{2}$Zr$_{2}$O$_{7}$ in a transverse magnetic field of strength $h=1$~T along the $[1,\bar{1},0]$ and $[1,1,1]$ directions. These measurements show that the $\mu$SR Knight shift at $h=1$~T takes the same form as the $h=0.1$~T susceptibility above $T \sim 1$~K, but below $T \sim 1$~K, both Knight shifts in $h=1$~T magnetic fields diverge from the magnetic susceptibility in $h=0.1$~T along $[1,\bar{1},0]$, with the Knight shifts turning down as a function of decreasing temperature and showing leveling off or plateau behaviour at the lowest temperatures measured.  

We then compare our DC magnetic susceptibility and Knight shift measurements to the bulk susceptibility calculated as $M/h$ (where $M$ is the magnetization) via fourth-order NLC calculations using the nearest-neighbour exchange parameters in the XYZ Hamiltonian [Eq.~\ref{eq:2}] estimated for Ce$_{2}$Zr$_{2}$O$_{7}$ in Refs.~\cite{Smith2022, Smith2023}, $(J_{\tilde{x}}, J_{\tilde{y}}, J_{\tilde{z}})$ = $(0.063, 0.062, 0.011)$~meV and $\theta = 0$. We show the calculations using the best-fit value of $g_z = 2.38$, according to fits of the bulk susceptibility measured from Ce$_{2}$Zr$_{2}$O$_{7}$ at $h = 0.1$~T, without using the measured Knight shift in the fitting process. We note that the best-fit value of $g_z$ is between the value of $g_z = 2.24$ estimated in Ref.~\cite{Smith2023} and the value of $g_z = 2.57$ corresponding to the pure $|m_J = \pm 3/2 \rangle$ ground state doublet estimated for Ce$_2$Zr$_2$O$_7$ in Refs.~\cite{Gaudet2019, Gao2019}. 

The lines in Fig.~\ref{Fig:6} show the NLC-calculated $M/h$ at $h=0.1$~T for each field direction above a low-temperature threshold of $T = 0.15$~K, as well as the $h=1$~T calculations above $T=0.01$~K for the $[1,\bar{1},0]$ field-direction and above $T = 0.05$~K for the $[1,1,1]$ field-direction, restricting each calculation to its region of convergence (within a small tolerance). As shown in Fig.~\ref{Fig:6}, our NLC calculations provide a remarkably good description of the measured magnetic susceptibility in a $h=0.1$~T field along the $[1,\bar{1},0]$ direction, as well as the Knight shift data for $h=1$~T fields along each direction. We note that our NLC calculations of $M/h$ using the previously-estimated values of $g_z$ given by $g_z$ = 2.24 (Ref.~\cite{Smith2023}) and $g_z = 2.57$ (Refs.~\cite{Gaudet2019, Gao2019}) also provide a good qualitative description of the measured data, and our fitting of $g_z$ using the $h = 0.1$~T susceptibility data was done to provide quantitative improvement only.  

The NLC calculations use the form of the XYZ nearest-neighbour Hamiltonian already established for Ce$_{2}$Zr$_{2}$O$_{7}$ (Refs.~\cite{Smith2022,Smith2023}) without further refinement of $(J_{\tilde{x}}, J_{\tilde{y}}, J_{\tilde{z}})$ or $\theta$. This description is very good, although not exact, possibly due to terms excluded from the XYZ Hamiltonian [Eq.~\ref{eq:2}] that may be significant at the low temperatures that are relevant here, such as weak further-than-nearest neighbour interactions for example, or differences between the bulk $M/h$ relevant to the calculations and the local $M/h$ probed by the Knight shift. For example, the NLC calculations show a reduced level of anisotropy compared to the measured data at $h = 1~T$, although importantly this splitting of $M/h$ between the $[1,\bar{1},0]$-field and $[1,1,1]$-field is in the same direction for both the calculations and the measured Knight shift. Nonetheless, this data and analysis establish that the difference in the form of the temperature dependence of $M/h$ at $h = 0.1$~T as measured in the bulk magnetic susceptibility of single crystal Ce$_{2}$Zr$_{2}$O$_{7}$, and $M/h$ at $h = 1$~T as measured in the $\mu$SR Knight shift of single crystal Ce$_{2}$Zr$_{2}$O$_{7}$, is largely magnetic field-induced, as opposed to a difference between bulk and local susceptibilities as is often the case in exotic magnets, due to the role of impurities. In turn, this reinforces the conclusion that impurities in our single crystal of Ce$_{2}$Zr$_{2}$O$_{7}$ play only a minor role, if any, in the spin dynamics of this quantum spin ice ground state material.  
  
\section{\label{sec:SummaryAndConclusions}Summary and Conclusions}

Our new $\mu$SR measurements on single crystals of the dipole-octupole quantum spin ice candidate Ce$_{2}$Zr$_{2}$O$_{7}$ in zero magnetic field show no signs of magnetic order or spin freezing and reveal a slower relaxation rate at low-temperature than that previously reported for powder Ce$_{2}$Zr$_{2}$O$_{7}$ in Ref.~\cite{Gao2019}, likely due to a lower sample oxidation in our carefully-prepared single crystal samples. The lack of significant decoupling in our longitudinal field $\mu$SR measurements provides clear evidence for the dynamic nature of the magnetic dipoles on the $\mu$s timescale in Ce$_{2}$Zr$_{2}$O$_{7}$ at $T = 0.1$~K. 

We find that the ZF and LF relaxation rate and stretching exponent are approximately constant below $T = 2$~K and $h = 2$~T, respectively, showing similar behaviour to that reported for powder Ce$_{2}$Sn$_{2}$O$_{7}$ (Ref.~\cite{Sibille2015}) which is known to have a low oxidation level.

For both the $[1,\bar{1},0]$ and $[1,1,1]$ field-directions, our new transverse field $\mu$SR measurements on single crystal Ce$_{2}$Zr$_{2}$O$_{7}$ in $h = 1$~T fields reveal Knight shifts whose temperature dependence tracks that of the DC magnetic susceptibility in a $h = 0.1$~T field along the $[1,\bar{1},0]$ direction, above $T \sim 1$~K.  Below $T\sim 1$~K, the temperature dependence of the Knight shifts in $h = 1$~T depart from the $\sim{\frac{1}{T}}$ temperature dependence of the magnetic susceptibility measured in a $h = 0.1$~T field applied along the $[1,\bar{1},0]$ direction, and the Knight shifts appear to saturate near $T =0.25$~K.     

Our fourth-order NLC calculations, using the nearest-neighbor XYZ Hamiltonian previously established for Ce$_{2}$Zr$_{2}$O$_{7}$ \cite{Smith2022, Smith2023}, provide a good description of both the DC magnetic susceptibility in a $h = 0.1$~T magnetic field and the Knight shift in a stronger $h = 1$~T magnetic field. The difference in the form of the temperature dependence at low temperature shown by the measured Knight shift, compared to the $\sim{\frac{1}{T}}$ form shown by the measured DC magnetic susceptibility, is thus ascribed to the field dependence of the magnetic susceptibility ($M/h$), as opposed to a difference between a bulk and a local susceptibility.  

This comprehensive $\mu$SR study of the DO pyrochlore Ce$_{2}$Zr$_{2}$O$_{7}$ clearly shows that the low energy spin dynamics of Ce$_{2}$Zr$_{2}$O$_{7}$ are sensitive to weak disorder within its quantum disordered ground state, even though this disorder is not sufficiently strong to induce either long-ranged order or a frozen spin state. This work also shows that NLC calculations employing earlier estimates for the microscopic XYZ Hamiltonian of Ce$_{2}$Zr$_{2}$O$_{7}$ can describe both the temperature dependence of the Knight shift and the DC magnetic susceptibility to the lowest temperature of measurement in each case, without further refinement of $(J_{\tilde{x}}, J_{\tilde{y}}, J_{\tilde{z}})$ and $\theta$, thereby providing corroboration of this earlier analysis~\cite{Smith2022, Smith2023}.

\begin{acknowledgments}

This work was supported by the Natural Sciences and Engineering Research Council of Canada (NSERC). We thank Owen Benton, Benedikt Placke, and Roderich Moessner at the Max Planck Institute for the Physics of Complex Systems for useful discussions and for their collaboration on related work. We thank Matthew Nugent for participation in $\mu$SR experiments included in this work. We greatly appreciate the technical support from Marek~Kiela and Jim~Garrett at the Brockhouse Institute for Materials Research, McMaster University. We thank Bassam Hitti, Donald Arseneau, and Gerald Morris for their support during measurements at TRIUMF. RS acknowledges the AFOSR Grant No. FA 9550-20-1-0235. The work of AAA and TJW was funded by the US Department of Energy, Office of Science, Office of Basic Energy Sciences.

\end{acknowledgments}

\section*{Appendix~A: Further Comparison of Relaxation Rates and Knight Shifts}

\begin{figure*}[t]
\linespread{1}
\par
\includegraphics[width=7.2in]{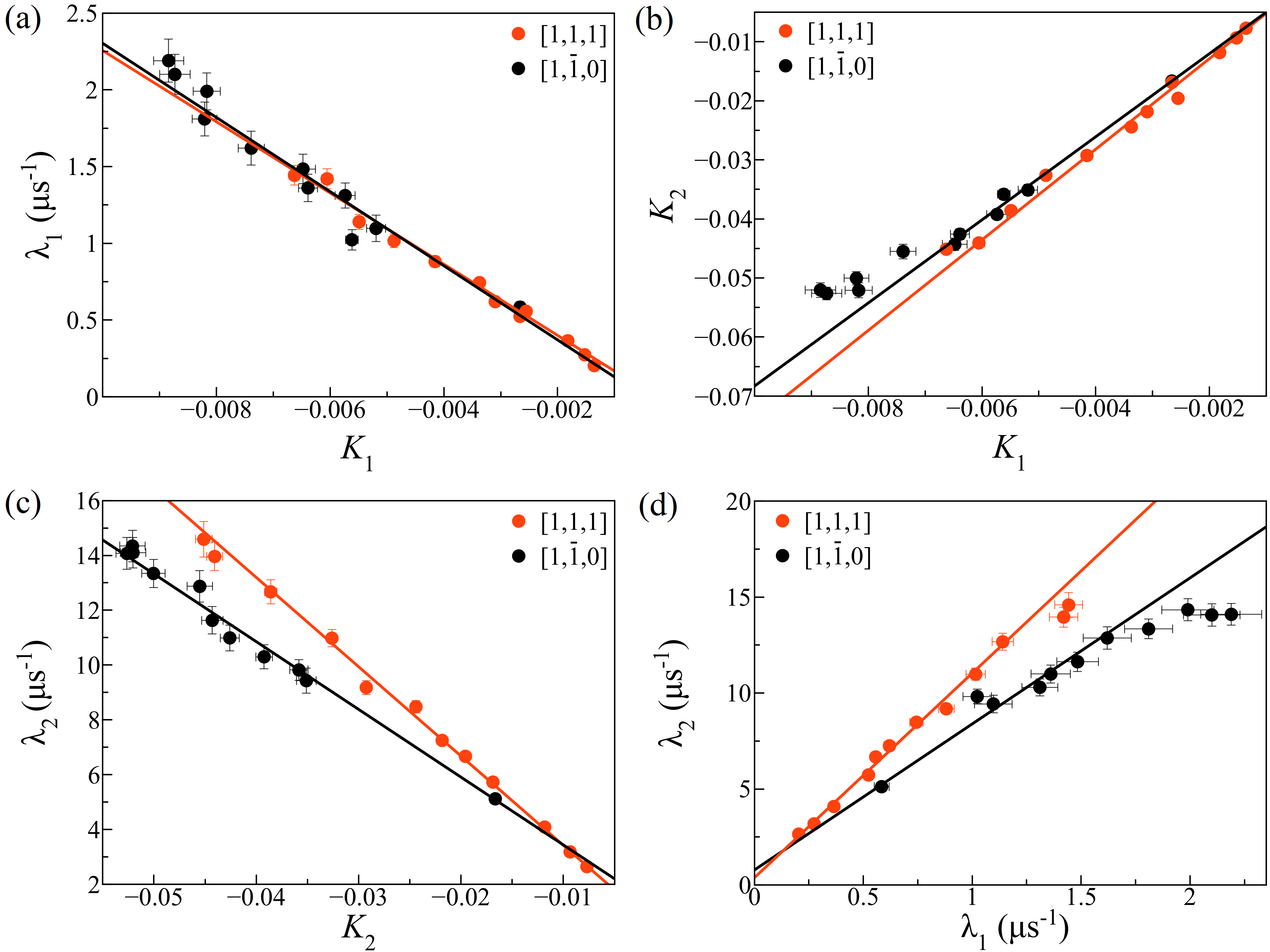}
\par
\caption{Comparing the measured relaxation rates ($\lambda_{1,2}$) and Knight shifts ($K_{1,2}$) from the two components of asymmetry measured from our single crystal samples of Ce$_{2}$Zr$_{2}$O$_{7}$ with a transverse magnetic field of strength $h=1$~T along the $[1,1,1]$ (red) and $[1,\bar{1},0]$ (black) directions. Specifically, we show comparisons of $K_1$ and $\lambda_1$ (a), $K_1$ and $K_2$ (b), $K_2$ and $\lambda_2$ (c), and $\lambda_1$ and $\lambda_2$ (d), and the best-fit line is shown for each comparison and field direction.} 
\label{Fig:7}
\end{figure*}

Fig.~\ref{Fig:7} shows the lines of best-fit to the linear relations displayed between the measured relaxation rates ($\lambda_{1,2}$) and Knight shifts ($K_{1,2}$) from our single crystal samples of Ce$_{2}$Zr$_{2}$O$_{7}$ in a $h=1$~T transverse magnetic field for both $[1,\bar{1},0]$ and $[1,1,1]$ field directions. For both the $[1,\bar{1},0]$ and $[1,1,1]$ field directions, the fact that these relations are linear implies that components 1 and 2 of the relaxation rates and Knight shifts detect the same magnetic behaviour but for different muon stopping sites.


\begin{table*}[h]
\label{TableI}
\begin{tabular}{|c|c|c|c|c|c|}
\hline
 & 1st Component & 1st Component & 2nd Component & 2nd Component \\
& $[1,\bar{1},0]$ Field & $[1,1,1]$ Field  & $[1,\bar{1},0]$ Field  & $[1,1,1]$ Field \\\hline

\begin{tabular}[c]{@{}c@{}}  \end{tabular}          
$K_0$ &-0.0087(2) & -0.0065(3) & -0.052(1) & -0.045(1) \\ \hline

\begin{tabular}[c]{@{}c@{}}  \end{tabular}          
$C$ & 0.0087(2) & 0.0062(3) & 0.057(1) & 0.046(1) \\ \hline

\begin{tabular}[c]{@{}c@{}}  \end{tabular}          
$\Delta$\,(K) & 0.81(5) & 0.90(8) & 1.04(4) & 0.98(4) \\ \hline

\end{tabular}
\caption{The best-fit values of $K_0$, $C$, and $\Delta$ from our fits of $K_0 + Ce^{-\Delta/T}$ to the different Knight shifts measured from our samples of Ce$_2$Zr$_2$O$_7$.}
\end{table*}
\clearpage
\bibliography{muSRStudyofCe2Zr2O7.bib} 
\end{document}